\documentclass{ws-procs9x6} 

\usepackage{graphicx,multirow,xspace,amsmath,amssymb}

\newcommand*{\kT}{\ensuremath{k_\mathrm{T}}\xspace}
\newcommand*{\pTjet}{\ensuremath{p_\mathrm{T}^{\mathrm{jet}}}\xspace}

\begin{document}
\title{Scaling properties of jet-momentum profiles with multiplicity}

\author{Antal G\'emes$^{a,b}$, R\'obert V\'ertesi$^{a}$, G\'abor Papp$^{c}$, Gergely G\'abor Barnaf\"oldi$^{a}$}

\address{$^a$Wigner Research Centre for Physics, MTA Centre of Excellence,\\ 
	29-33 Konkoly-Thege Mikl\'os Str., 1121 Budapest, Hungary}
\address{$^b$Trinity College, University of Cambridge,\\Cambridge CB2 1TQ, United Kingdom}
\address{$^c$Institute of Physics, E\"otv\"os Lor\'and University,\\P\'azm\'any P\'eter s\'et\'any 1/A, 1117 Budapest, Hungary}

\begin{abstract}
We study the structure of jets in proton-proton collisions at LHC energies using \textsc{Pythia} 8 Monte Carlo simulations. 
We demonstrate that the radial jet profiles exhibit scaling properties with charged-hadron event multiplicity over a broad transverse-momentum range.
We also provide parametrizations of the jet profiles based on different statistically-motivated analytical distributions.
Based on this we propose that the scaling behavior stems from fundamental statistical properties of jet fragmentation.

\end{abstract}


\bodymatter

\section{Introduction}\label{sec:intro}

Measurements of jet profiles in high-energy collisions are sensitive probes of QCD parton splitting and showering~\cite{Seymour:1997kj,Ellis:1992qq,Vitev:2008rz}. Precise understanding of the jet structures are essential for setting the baseline not only for nuclear modification of jets in heavy-ion collisions~\cite{Vitev:2008rz}, but also for possible cold QCD effects that may modify jets in high-multiplicity proton-proton collisions~\cite{Varga:2018isd,Varga:2019rhi}.

Although multiplicity-differential jet-shape analyses are feasible up to high transverse momenta (\pTjet) by the current precision of the large LHC experiments, currently available experimental results are either multiplicity-inclusive or transverse-momentum inclusive~\cite{Chatrchyan:2012mec,Chatrchyan:2013ala}. 
In simulation studies, however, the multiplicity and \pTjet-dependence of differential and integrated jet structure observables were explored in details, using several models implemented in the \textsc{Pythia} 8 as well as the \textsc{Hijing}++ event generators~\cite{Sjostrand:2014zea,Barnafoldi:2017jiz}. These simulations are consistent with data wherever available, but also provide predictions for the jet-structure observables in several multiplicity classes over a wide transverse-momentum range. In these results, the jet-momentum density was found to be independent of multiplicity at a given radius $R_\mathrm{fix}$, which was in turn insensitive to the choice of event generators and tunes, parton distribution function sets, the presence and modelling of semi-hard processes (such as multiple-parton interactions, color-reconnection or minijet production), and even of jet clustering algorithms~\cite{Varga:2018isd,Varga:2019rhi}. 

In the current study we demonstrate that the radial jet-momentum profile exhibits a scaling behavior with charged-hadron event multiplicity. We also show that the jet-momentum profiles can be adequately parametrized based on different statistically-motivated analytical distributions. This suggests that the scaling behavior stems from fundamental statistical properties of jet fragmentation, independently of the above-mentioned choices of modelling and reconstruction settings.

\section{Analysis method}\label{sec:ana}

Differential jet structures were simulated following the procedure described elsewhere~\cite{Varga:2018isd}. We used a sample of 5 million proton-proton collisions at $\sqrt{s}=7$ TeV collision energy, obtained using \textsc{Pythia} 8 (version 8.226) with the tune 4C~\cite{Corke:2010yf}. We reconstructed jets using the anti-$\kT$ clustering algorithm~\cite{Cacciari:2008gp} with a resolution parameter $R=0.7$ in the mid-rapidity range $|\eta|<1$. The radial transverse momentum distribution inside the jet cone, also called the differential jet shape variable, is defined as
\begin{equation}
\varrho(r) = \frac{1}{\delta r} \frac{1}{\pTjet}\sum\limits_{\substack{r_a < r_i < r_b}} p_{\mathrm{T}}^i \,,
\label{eqn:DiffJet}
\end{equation}
where $p_{\mathrm{T}}^i$ is the transverse momentum of a particle inside a ring with a thickness of $\delta r$ and an inner radius $r_a = r - \frac{\delta r}{2}$ and outer radius $r_b = r + \frac{\delta r}{2}$ around the jet axis and \pTjet is the transverse momentum of the whole jet. The distance of a given particle $i$ from the jet axis can be expressed in terms of the azimuthal angle $\varphi_i$ and the pseudorapidity $\eta_i$ as $r_i = \sqrt{(\varphi_i - \varphi_\mathrm{jet})^2 + (\eta_i - \eta_\mathrm{jet})^2}$.
In the current analysis we used a binning of $\delta r$=0.05 and restricted ourselves to the fiducial region $r<0.6$ to avoid edge effects. We categorized the events based on charged-hadron event multiplicity within the mid-rapidity region $|\eta|<1$. 

In the following we will use two statistically-motivated ans\"atze to describe the radial jet-momentum distributions~\cite{Wilk:2012zn,Biro:2014yoa},
\begin{equation}
    \varrho(r) = C \frac{\Gamma(rk+a)}{\Gamma(a) \Gamma(rk+1)} p^{rk} (1-p)^{a}\,,
    \label{eqn:NBD}
\end{equation}
the negative binomial distribution (NBD) function, inspired by an analogy with the Koba--Nielsen--Olesen (KNO) scaling~\cite{Koba:1972ng,Hegyi:2000sp}, and
\begin{equation}
    \varrho(r) = C r^\gamma e^{-\alpha r}\,, 
    \label{eqn:Poisson}
\end{equation}
a simpler parametrization based on the Poisson distribution, a special case of the negative binomial distribution~\cite{NBD}. These function forms were fit to the data together with a $\varrho_{\rm bg}(r) = br$ linear term representing a uniform background under the jet. Thereafter the background was subtracted from the $\varrho(r)$ distributions separately for each \pTjet and multiplicity range.

\section{Results}\label{sec:results}

\begin{figure}
\centering
\includegraphics[width=0.5\textwidth]{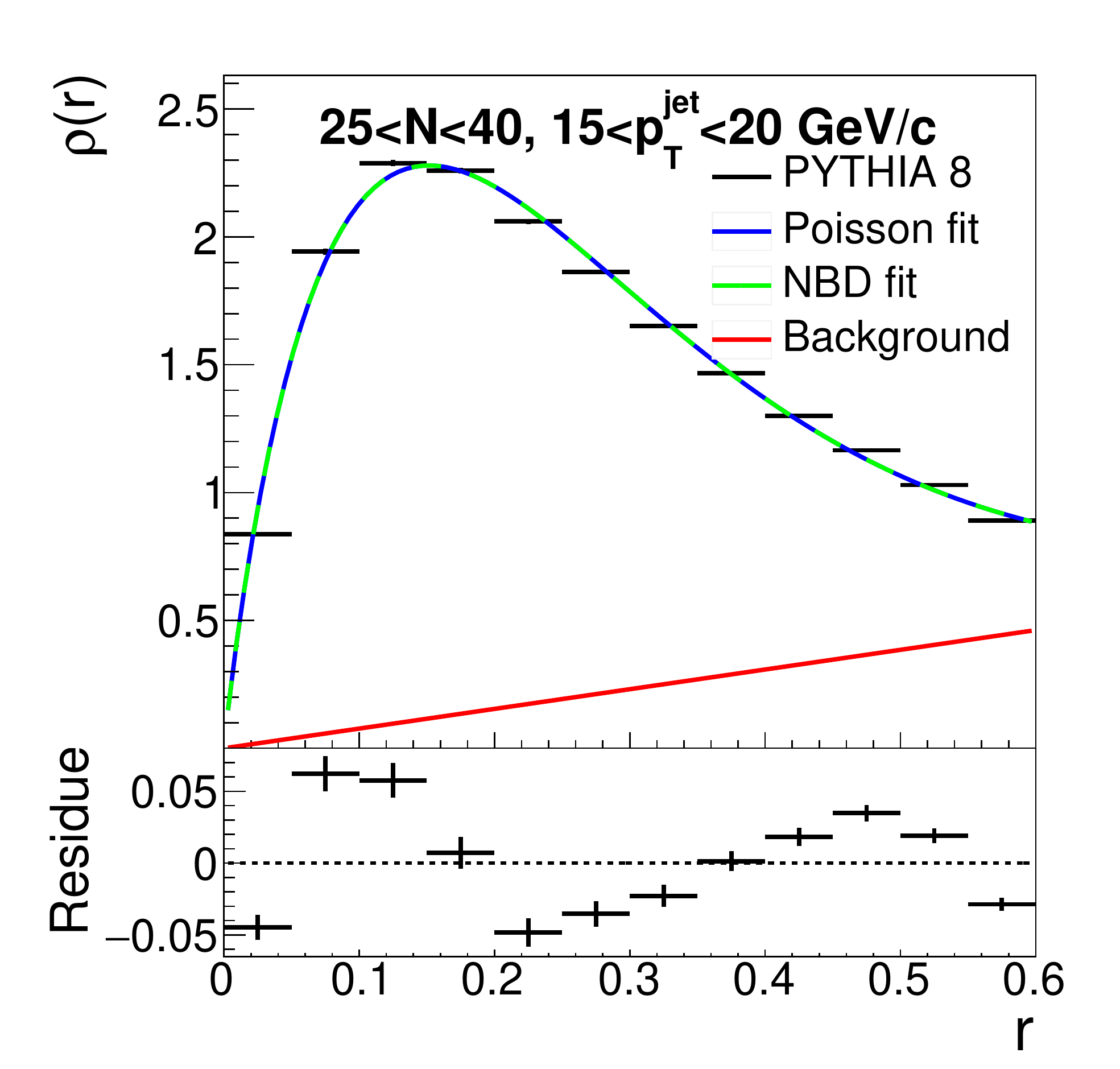}%
\includegraphics[width=0.5\textwidth]{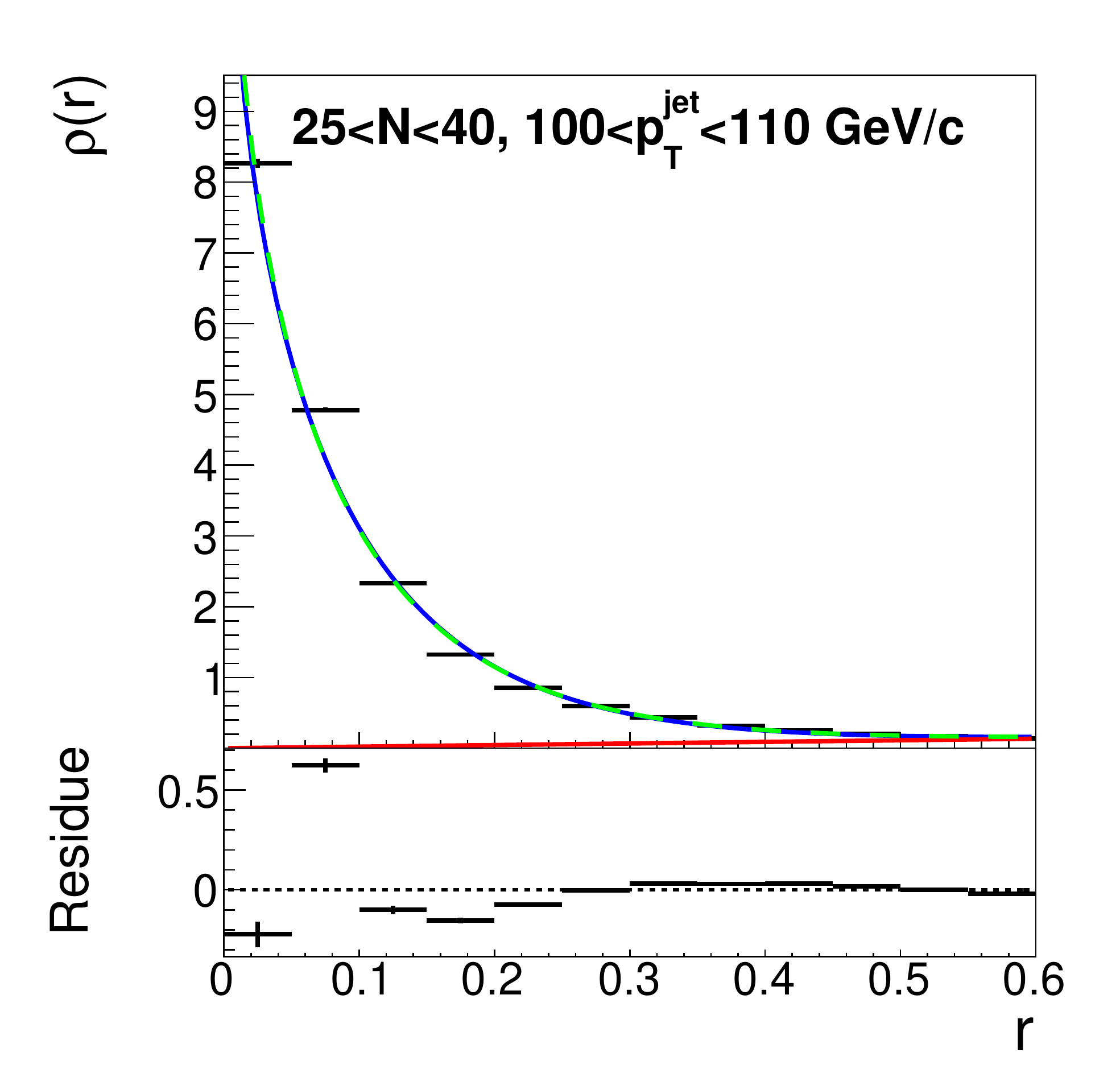}
\includegraphics[width=0.5\textwidth]{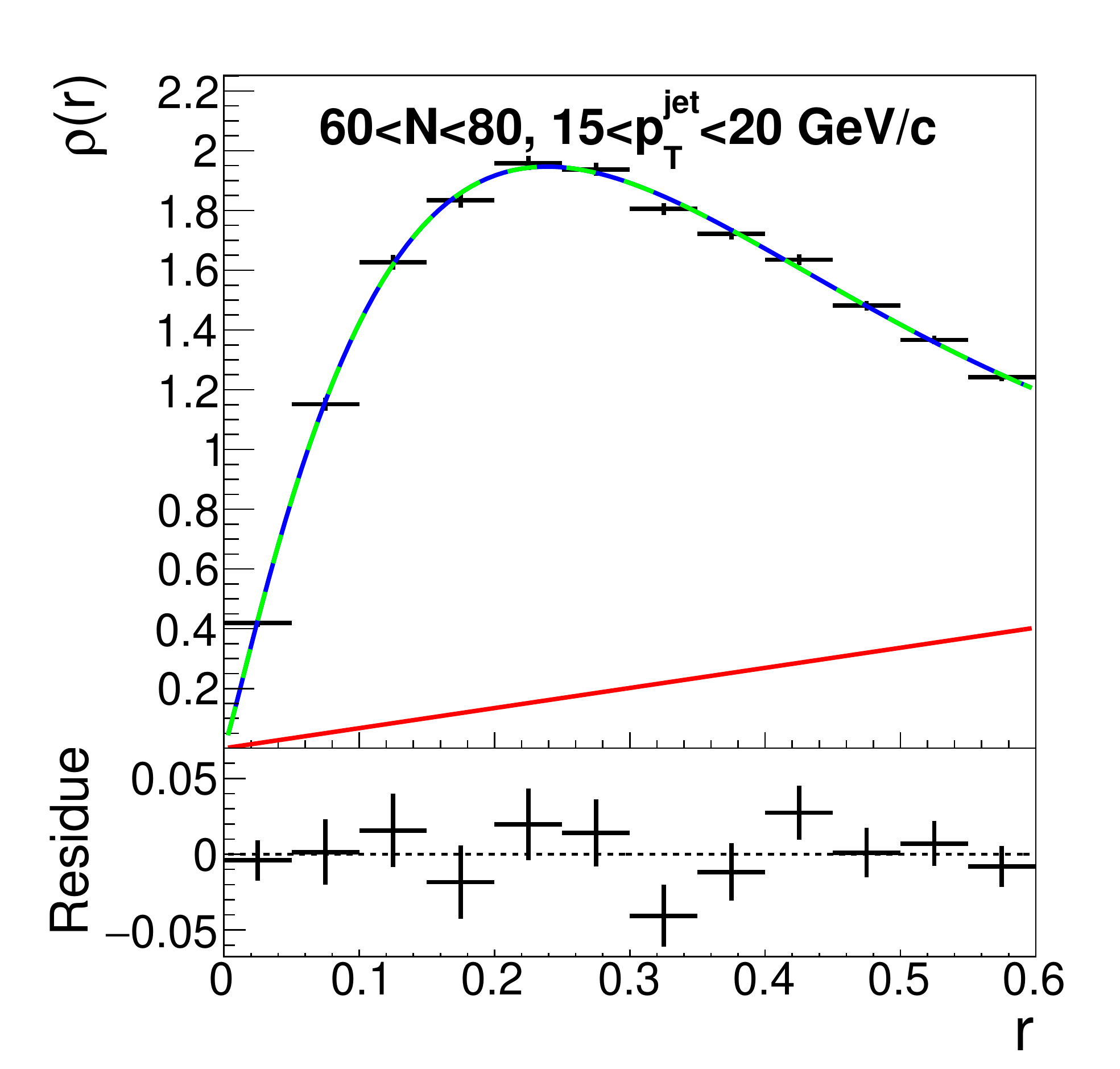}%
\includegraphics[width=0.5\textwidth]{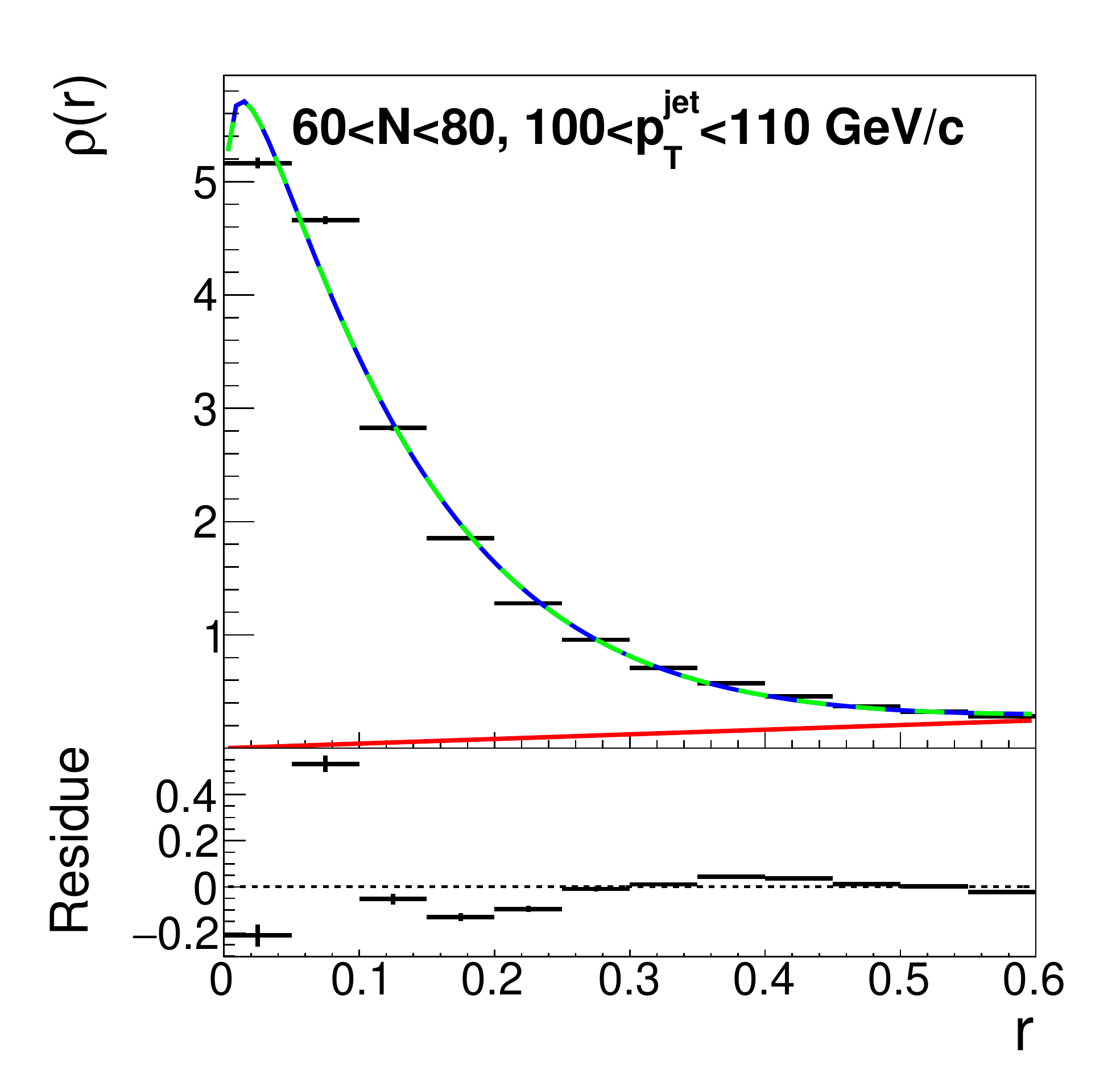}
\caption{The Poisson distribution and NBD fits over a linear background in the $25<N<40$ and $60<N<80$ multiplicity bins (top panels in the upper left and right subfigures, respectively) as well as the $15<\pTjet<20$ GeV/$c$ and $100<\pTjet<110$ GeV/$c$ jet transverse momentum ranges (top panels in the lower left and right subfigures, respectively). The lower panels of each subfigure show the residue after subtracting the Poissonian parametrization from the simulations.}
\label{fig:fits}
\end{figure}

Figure \ref{fig:fits} shows the results from the simultaneous fits of the Poissonian form together with a linear background, as well as the fit of the NBD to the differential jet shapes, also with a linear background.
The overall shape of the jets is described well by both the Poissonian and the NBD parametrizations, with virtually no difference between the two cases. There is a slight difference, however, in the peak region between the fits and the data in some cases (in the order of 10\%). It is also to be noted that the peak structure is not resolved in case of the narrow jets at high \pTjet and low multiplicities, and therefore the model description is underdetermined in such cases. 

\begin{figure}
\begin{center}
\includegraphics[width=0.999\textwidth]{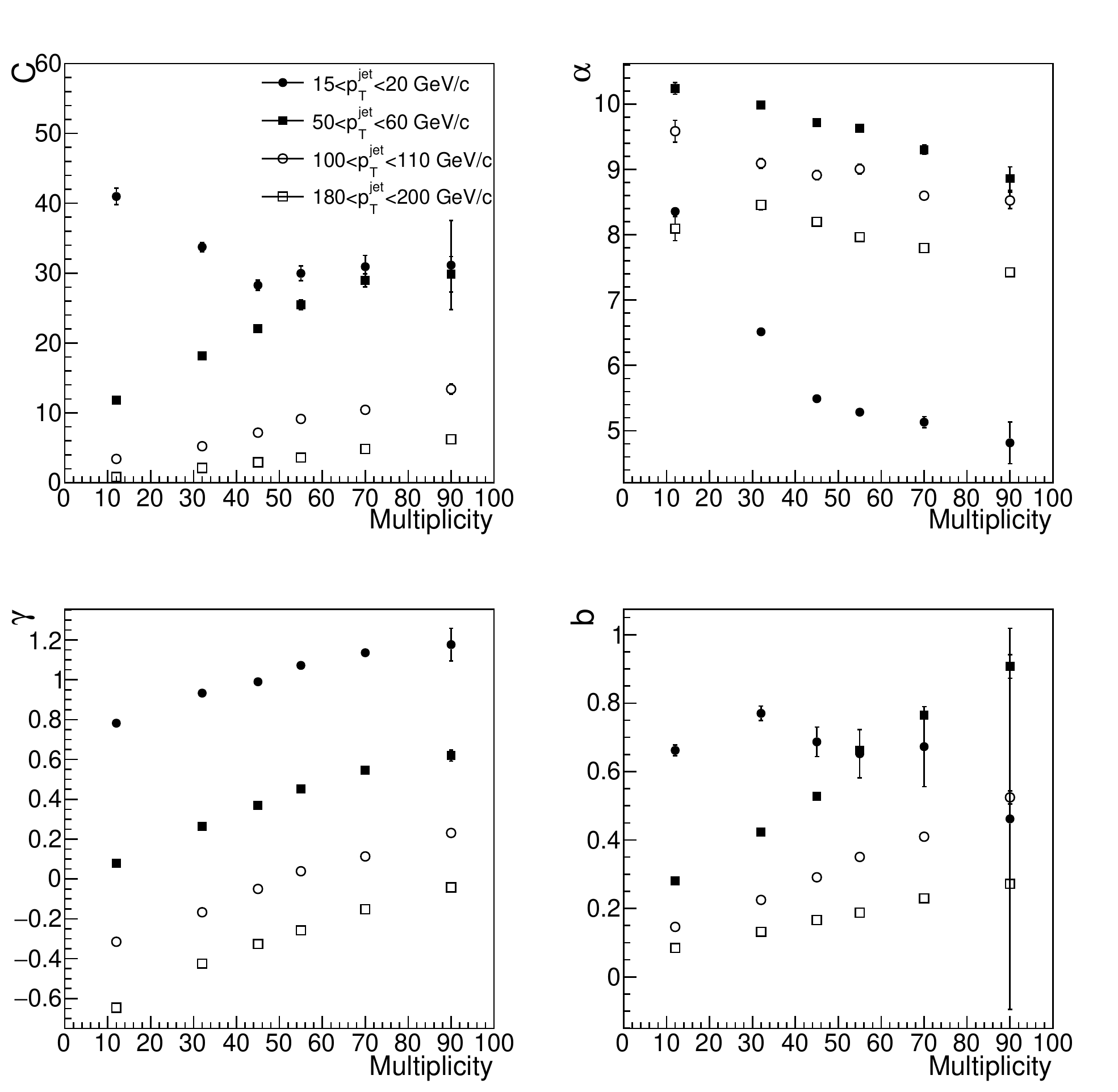}
\end{center}
\caption{Parameters of the Poisson distributions ($C$, $\alpha$, $\gamma$) describing the jet profiles and the background ($b$) in the $15<\pTjet<20$ GeV/$c$, $50<\pTjet<60$ GeV/$c$, $100<\pTjet<110$ GeV/$c$ and $180<\pTjet<200$ GeV/$c$ jet transverse momentum ranges.}
\label{fig:FitParameters}
\end{figure}

When using the negative binomial distribution to fit the jet-momentum profiles, the parameter $p$ of Eq.~\eqref{eqn:NBD} converged to either 1 (typically for wide jets, with high multiplicity and/or low \pTjet), or to 0 (for narrow jets). In case of $p \xrightarrow{}1$, NBD reduces to a Poisson distribution~\cite{NBD}. When the parameter $p$ converged to 0, the parameter $a$ also converged to 0, which is another limiting case for the Poisson distribution with $\gamma \xrightarrow{} -1$. Since all negative binomial distribution fits either reduced to a Poisson distribution or a limiting case of it, and the fit results are very similar to those obtained from Eq.~\eqref{eqn:Poisson}, the Poissonian ansatz was used in the followings. The fit parameter values of the Poissonian parametrization with the background can be seen in Fig.~\ref{fig:FitParameters} in function of event multiplicity for four selected \pTjet ranges.

We found that the jet-momentum profiles exhibit a scaling property of the form
\begin{equation}
    \varrho_N(r) = \lambda(N) f \left( \frac{r}{\kappa(N)} \right)\,,
\end{equation}
where $N$ is the charged-hadron event multiplicity. 
Since the differential jet shape variable defined in Eq.~\eqref{eqn:DiffJet} is normalized to unity, one would expect the product $\lambda \kappa$ to be a constant. However, this does not hold for two reasons. First, for wide jets, a significant fraction of the jet is outside the $r<0.7$ cone where the normalization was preformed, and second, the data was normalized when it still contained the background. The exclusion of the tail of the momentum profiles leads to an increase of $\lambda$ with the increase of multiplicity, while the inclusion of the background before the normalization decreases $\lambda$. Since the background increases with the multiplicity, this leads to a decrease of $\lambda$ with multiplicity. For these reasons $\lambda$ and $\kappa$ were treated independently in the scaling. 

In each \pTjet range we used a certain multiplicity bin as a reference, while the data from other multiplicity bins were matched onto it by scaling $\lambda$ and $\kappa$ of the fit function of the reference using $\chi^2$ minimization. The parameters $\lambda$ and $\kappa$ are defined so that they are unity for the lowest multiplicity bin.
At low \pTjet, low-multiplicity bins were chosen as reference since high-multiplicity bins lack sufficient statistics. In higher \pTjet ranges, distributions in low-multiplicity bins are very narrow and thus the fits are underdetermined, while the wider jets at high multiplicity lack the information about the tail of the momentum distribution. Therefore intermediate-multiplicity bins were used as references in higher \pTjet ranges.

\begin{figure}
\begin{center}
\includegraphics[width=0.999\textwidth]{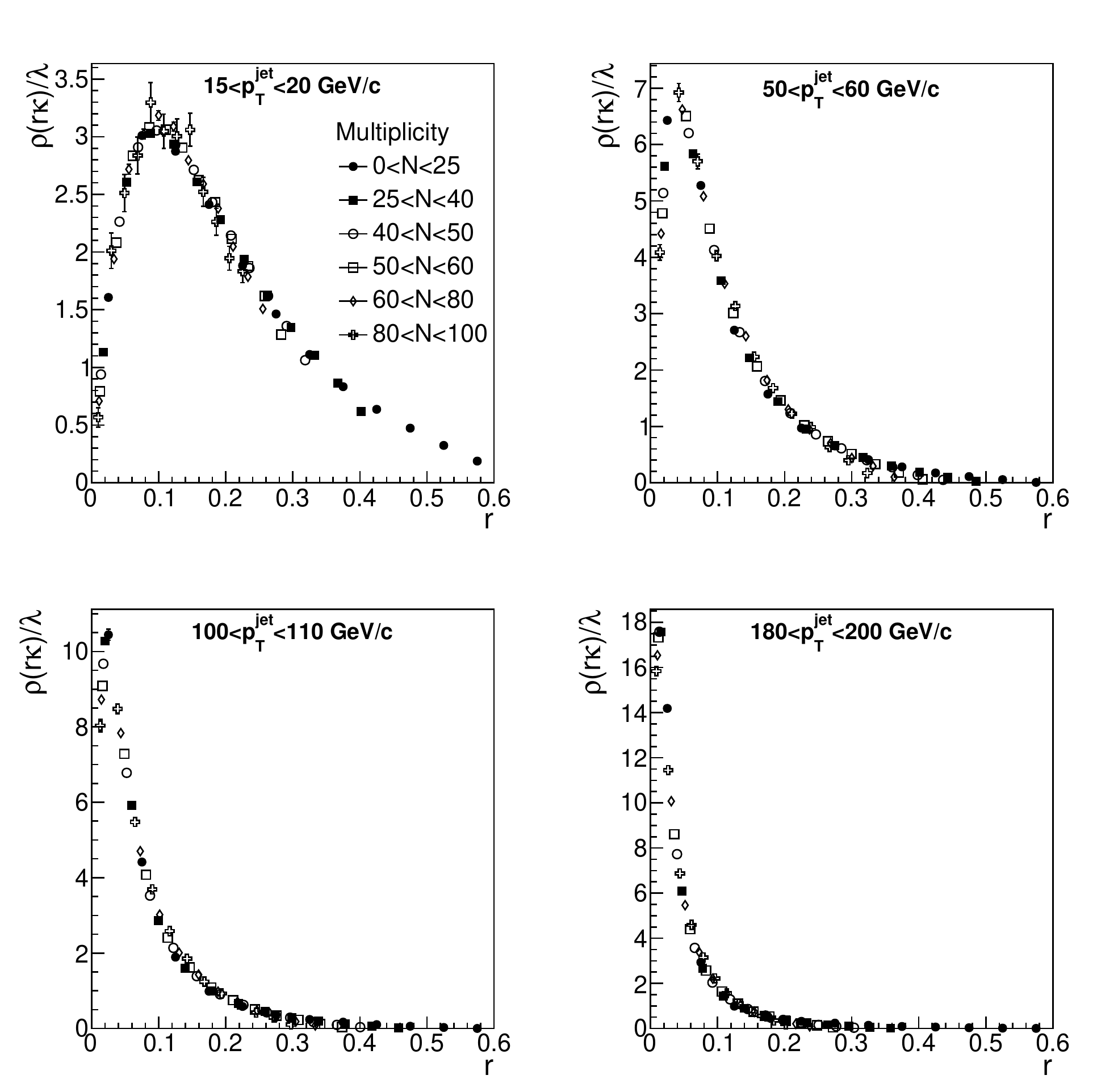}
\end{center}
\caption{The background-subtracted scaled jet-momentum profiles in four representative \pTjet ranges. All multiplicity bins are rescaled onto the lowest multiplicity bin.}
\label{fig:scaled}
\end{figure}

\begin{figure}
\begin{center}
\includegraphics[width=0.999\textwidth]{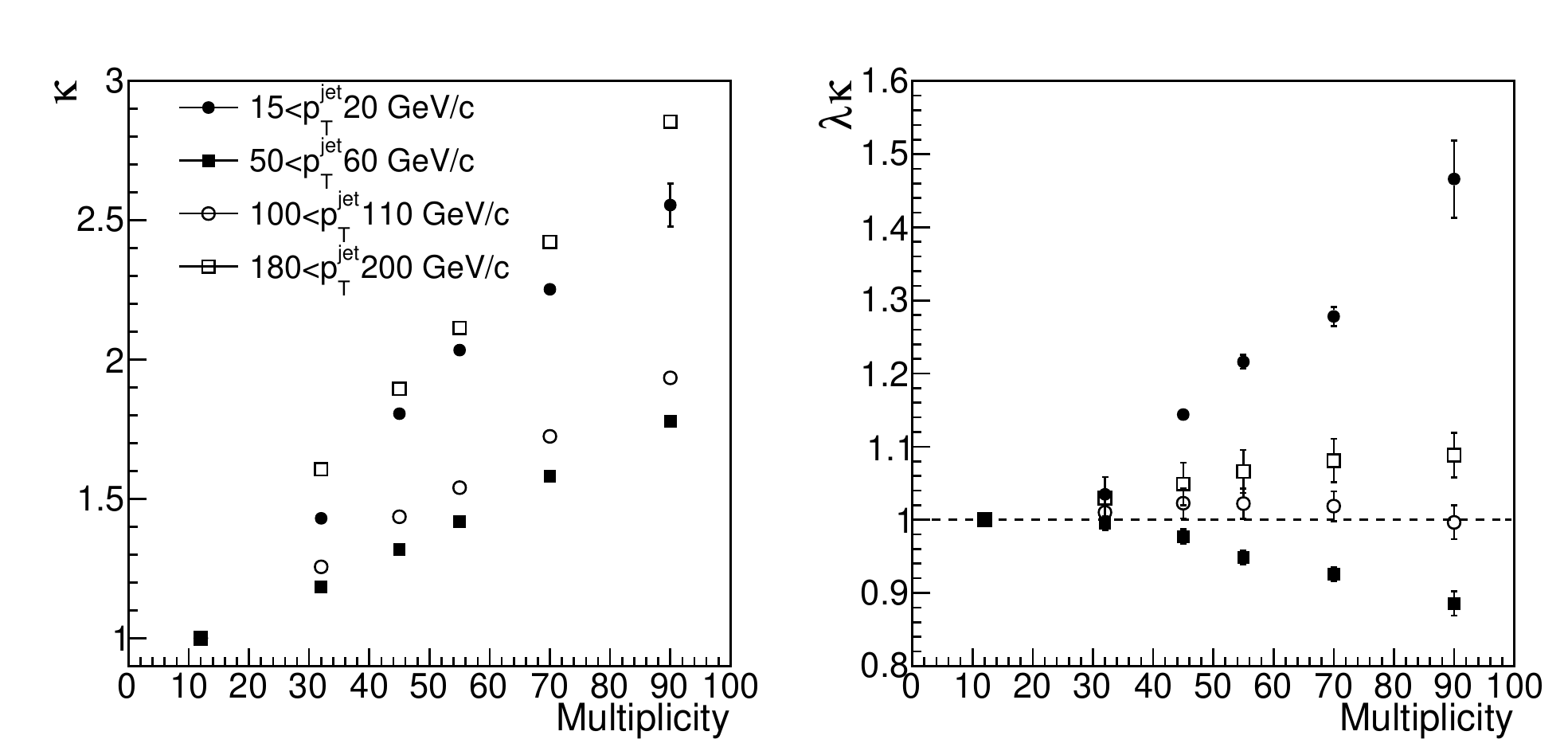}
\end{center}
\caption{The scaling parameter $\kappa$ (left) and the product $\lambda \kappa$ (right) in four different \pTjet ranges as a function of multiplicity, for four selected \pTjet ranges. The $\kappa$ and $\lambda$ parameters are calculated relative to the lowest multiplicity bin.}
\label{fig:ScalingPar}
\end{figure}

The background-subtracted scaled jet-momentum profiles are shown in Fig.~\ref{fig:scaled} for four selected \pTjet ranges.
A limitation of our method is that the fit function forms used for the scaling do not provide a perfect description in the statistical sense. Nevertheless, the scaling works adequately well, to the level of a 10\% deviation in the worst case, at low \pTjet and extreme multiplicities where the influence of the background is the highest. 
We show the values of the scaling parameter $\kappa$, as well as the product of the two scaling parameters,  $\lambda \kappa$, in Fig.~\ref{fig:ScalingPar}.
Even though we do not expect $\lambda \kappa$ to be at unity, for all \pTjet ranges but the lowest one the deviation from it is less than 10\%.

\begin{figure}
\begin{center}
\includegraphics[width=0.999\textwidth]{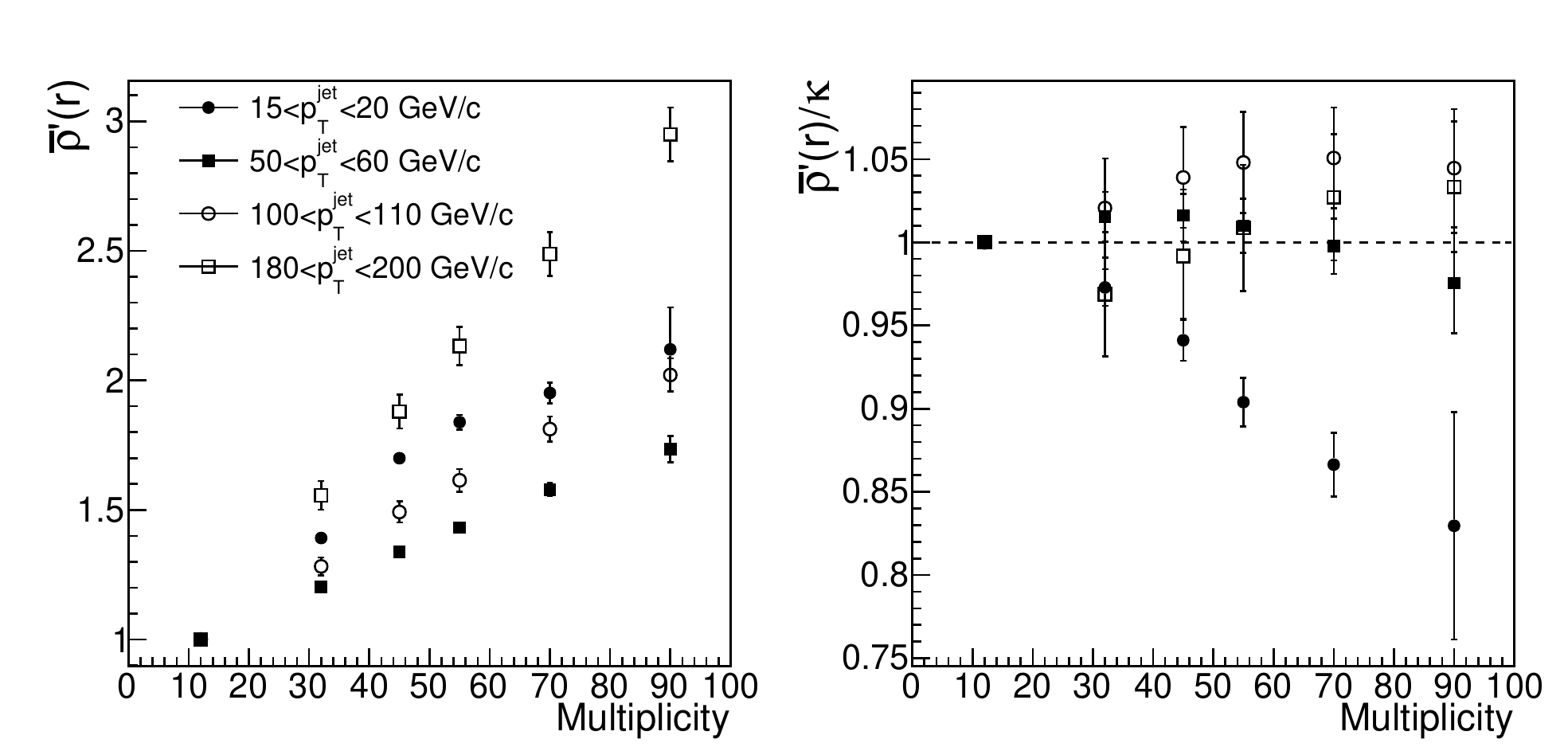}
\end{center}
\caption{The rescaled mean $\overline{\varrho}^\prime$ of the Poisson distributions (left), and $\overline{\varrho}^\prime/\kappa$ (right) in function of multiplicity, for four selected \pTjet ranges.}
\label{fig:ScalingMeans}
\end{figure}

Assuming that the chosen function shape adequately describes the data and the scaling holds, the means of the Poisson distributions could also be used to scale the multiplicity bins onto each other. From Eq.~\eqref{eqn:Poisson} the mean of the Poisson distribution can be expressed as 
\begin{equation}
 \overline{\varrho}=(\gamma+1)/\alpha  \,. 
\end{equation}
Since $\kappa$ is determined relative to the lowest multiplicity bin, we also rescale the means to its value at the lowest multiplicity bin in the case of each \pTjet range. 
Fig.~\ref{fig:ScalingMeans} shows this rescaled mean, $\overline{\varrho}^\prime$ as well as its ratio to the $\kappa$ scaling parameter. Except for the lowest \pTjet range, $\kappa/{\overline{\varrho}^\prime}$ is within 5\% from unity. This further supports that the radial jet-momentum profiles exhibit a scaling behavior with multiplicity, and that the Poisson distribution is an adequate description of the data to find the parameters of the scaling.

\section{Conclusion}\label{sec:concl}

We demonstrated that background-subtracted radial jet-momentum profiles from simulated proton-proton collisions at LHC energies exhibit a scaling behavior with event multiplicity taken at mid-rapidity. We provided a parametrization of the momentum profiles with a Poisson distribution and showed that an assumed negative binomial distribution also reduces to a Poissonian. This suggests that the scaling behavior stems from fundamental statistical properties of jet fragmentation, independently of the choice of particular model settings. Multiplicity-dependent measurements in a broad momentum range would be essential to cross-check these findings in real data.

\section*{Acknowledgement}
This work was supported by the Hungarian National Research, Development and Innovation Office (NKFIH) under the contract numbers OTKA K120660, K123815, FK131979 and the NKFIH grants 2019-2.1.11-T\'ET-2019-00078, 2019-2.1.11-T\'ET-2019-00050, 2019-2.1.6-NEMZ\_KI-2019-00011 and 2020-1.2.1-GYAK-2020-00013, as well as the THOR Cost Action CA15213. The authors also acknowledge the computational resources provided by the Wigner GPU Laboratory.

\end{document}